\documentclass[english,aps,prper,longbibliography,reprint,superscriptaddress]{revtex4-1}   
\usepackage{lipsum}
\usepackage[T1]{fontenc}	
\usepackage[latin9]{inputenc}	
\usepackage{geometry}		
\usepackage{graphicx}
\usepackage{color}
\usepackage[above,below]{placeins}	
\usepackage{times}
\usepackage[colorlinks=true,allcolors=blue]{hyperref}
\usepackage{multirow}
\usepackage{rotate}
\usepackage{lineno,xcolor}
\usepackage[colorlinks=true,allcolors=blue]{hyperref}
\usepackage{multirow}
\usepackage{rotate}

\usepackage{amsmath,amssymb}
\usepackage{array}
\usepackage[caption = false]{subfig}

\begin{document}

\title{Online test administration results in students selecting more responses to multiple-choice-multiple-response items}

\author{Alexis Olsho}
\affiliation{Department of Physics and Meteorology, United States Air Force Academy, 2354 Fairchild Drive, USAF Academy, CO 80840 USA}

\author{Trevor I.\ Smith}
\affiliation{Department of Physics \& Astronomy and Department of STEAM Education, Rowan University, 201 Mullica Hill Rd., Glassboro, NJ 08028, USA}

\author{Philip Eaton}
\affiliation{School of Natural Sciences and Mathematics, Stockton University, Galloway, NJ 08205, USA}

\author{Charlotte Zimmerman}
\affiliation{Department of Physics, University of Washington, Box 351560, Seattle, WA 98195-1560, USA}

\author{Andrew Boudreaux}
\affiliation{Department of Physics \& Astronomy, Western Washington University, 516 High St., Bellingham, WA 98225, USA}

\author{Suzanne White Brahmia}
\affiliation{Department of Physics, University of Washington, Box 351560, Seattle, WA 98195-1560, USA}

\begin{abstract}

Multiple-choice multiple-response (MCMR) items (i.e., multiple-choice questions for which more than one response may be selected) can be a valuable tool for assessment. Like traditional multiple-choice single-response questions, they are easy to score, but MCMR items can provide more information about student thinking by probing multiple reasoning facets in a single problem context. In this paper, we discuss differences in performance on MCMR items that result from differences in administration method (paper vs. online). In particular, we find a tendency for ``clickiness'' in online administration: students choose more responses to MCMR items when taking the electronic version of the assessment. Based on these findings, we recommend that online administration be compared only to other online administrations and not to paper administrations when comparing student results on MCMR items. These results also suggest that MCMR items provide a unique opportunity to probe differences in online and on-paper administration of low-stakes assessments.

\end{abstract}

\maketitle

\section{Introduction and Background}

Research-based assessments (RBAs) are now widely used to examine student understanding and learning in physics instruction. Best practices are established for the administration of RBAs both in-person (on paper or electronically) and as an out-of-class, online activity \cite{madsen2017,madsenphysport}. While research suggests that RBAs originally designed to be given in-person can generally be administered online without affecting student performance \cite{bonham2008,nissen2018,wilcox2014}, researchers also recommend that instruments are validated separately for online, unproctored use \cite{bonham2008}. 

We have developed the Physics Inventory of Quantitative Literacy (PIQL) to assess students' quantitative reasoning in introductory physics contexts \cite{white2021}. The PIQL includes several ``multiple-choice-multiple-response'' (MCMR) items. MCMR items are multiple-choice items for which there may be more than one correct response, and for which students are encouraged to select all answers that apply. MCMR items may be a useful tool for probing multiple facets of student understanding, providing more insight into student reasoning than standard single-response multiple-choice items in situations where free-response questions are not feasible \cite{wilcox2015}. 

An example of an MCMR item from the PIQL is shown in Fig.~\ref{fig:workProb}. This item is intended to probe multiple facets of mechanical work as a signed quantity. The correct answers are d and g---choice d relates to a more mathematical understanding of the calculation and interpretation of a negative scalar product, while answer choice g relates to a interpretation of negative net work as causing a decrease in mechanical energy of a system. 

\begin{figure}[tb]
\framebox{\parbox{0.45\textwidth}{\raggedright
A hand exerts a constant, horizontal force on a block as the block moves along a frictionless, horizontal surface. No other objects do work on the block. For a particular interval of the motion, the hand does $W= -2.7$~J of work on the block. Recall that for a constant force, $W = \vec{F}\cdot\Delta\vec{s}$.\\

\vspace{1em}

\noindent
Consider the following statements about this situation.  Select the statement(s) that \textbf{must be true.}\\ \textit{\textbf{Choose all that apply.}}

\vspace{1em}

\begin{itemize}
\item[a.] The work done by the hand is in the negative direction. 
\item[b.] The force exerted by the hand is in the negative direction. 
\item[c.] The displacement of the block is in the negative direction. 
\item[d.] The force exerted by the hand is in the direction \textit{opposite to} the block's displacement. 
\item[e.] The force exerted by the hand is in the direction \textit{parallel to} the block's displacement. 
\item[f.] Energy was added to the block system. 
\item[g.] Energy was taken away from the block system. 
\end{itemize}

}}

    \caption{PIQL MCMR item that probes understanding of the negative sign in the context of mechanical work. The correct responses are d and g.}
    
    \label{fig:workProb}
\end{figure}

The PIQL was originally developed as an in-person, proctored assessment, but shifted to online administration in early 2020 due to the COVID-19 pandemic. Though the shift was not planned, online research-based assessments offer important affordances associated with easing the logistics of test administration \cite{olsho2020perc}. We therefore began to collect evidence of the PIQL's validity as an online assessment, with an eye toward dissemination of the PIQL as an online instrument for widespread use. While preliminary research revealed that student performance on the PIQL was largely unaffected by administration method, we found significant differences in the number of responses provided by students on the PIQL's MCMR items \cite{olsho2020perc}.

Despite the potential for assessing partial or incomplete student reasoning MCMR items might provide, little research has been done comparing student test-taking behavior based on administration format. Wilcox and Pollock report that student performance on coupled multiple-response items on the Colorado upper-division electrostatics (CUE) diagnostic was similar across online and in-person administration methods, but an examination of possible differences was not a focus of their work \cite{wilcox2015}. We also note that the coupled multiple-response items on the CUE are a particular style of MCMR item, and the CUE diagnostic (as its name suggests) is used with upper-division students. It is possible that their findings do not apply to all MCMR items, or all students enrolled in college-level physics courses.

In this paper, we explore factors that may affect introductory physics students' performance on the PIQL's MCMR items. In particular, we investigate whether administration method affects students' response patterns for MCMR questions. 
We use our findings to discuss insights into how students interact with online, low-stakes assessments and implications for item and inventory development. 

\section{Methods}

Most of the data discussed in this paper were collected from students enrolled in the first quarter of the calculus-based introductory physics sequence at a large public research university in the Western US (``Institution 1''). 

At Institution 1, data were collected over eight academic terms. In four of these, the PIQL was administered in a course recitation session proctored by a TA. Students ($N = 1697$) read items from a 5-page stapled and laminated packet and recorded their responses on a paper answer form as well as electronically, where the paper copy served as a backup if the electronic submission failed. For each of these in-person administrations of the PIQL, the MCMR items were interspersed with single-response (SR) items throughout the test. Before starting, the students were informed of the purpose of the PIQL by the TA. Students were reminded in multiple ways that they could choose more than one response on MCMR items. The proctor reminded students verbally. Students were reminded in writing at the top of each page of the packet that contained an MCMR item.  Finally, students were prompted to "choose all that apply" in the question stem for each MCMR item.  There was no official time limit, though students were expected to hand in their work by the end of the 50-minute recitation session. In our experience, students had little difficulty completing the PIQL in this time, with most finishing within 40 minutes. 

In the other four terms at Institution 1 ($N = 1404$), the PIQL was administered unproctored and entirely online using the University's existing survey/quiz platform. When administered online, the PIQL had a 50-minute time limit. Each item was shown in a browser window on its own; students were not able to backtrack in the PIQL \footnote{We recognize that not being able to look at questions already completed is a significant change from in-class practices, but deemed it necessary for test security purposes, to decrease the likelihood that test items were copied by students.} and were not shown a summary of their work.

With one exception (described below), all of the MCMR items were moved to the end of the PIQL for online administration. After answering the last SR item, students saw a page with no instrument item, but rather a statement that the remaining questions on the survey might have more than one correct response, and that students should choose all answers that they feel are correct. At the top of the page for each of the remaining items (all MCMR), students saw a reminder that the question might have more than one correct response. As in the in-person administration, the question stem for each MCMR item prompted students to ``choose all that apply.''

Additional data were collected from students enrolled in the first semester of introductory calculus-based physics at a US military academy (``Institution 2''). At Institution 2, the PIQL was administered in-person during a regular class period and proctored by the course instructor. Students ($N = 282$) read items from a 5-page stapled and laminated packet and recorded their responses on a bubble sheet. MCMR items were grouped at the end of PIQL, with a reminder at the top of each page with MCMR items (the last two pages of the test).

While our primary focus for this research is to find the effect of on-paper vs. online administration on MCMR performance and answer choices, we recognize that changing item order (i.e., moving all of the MCMR items to the end of the assessment) may also have an effect on student response patterns. To assess the effect of question order on the number of responses chosen, we presented the PIQL online to a class of students ($N=83$) enrolled in the first quarter of the calculus-based introductory physics sequence at Insititution 1. Approximately half the students ($N=43$) saw the MCMR items grouped at the end of the PIQL, while the remaining 40 students saw the MCMR items interspersed with the SR items.

\section{Results}

\begin{figure}
 
    \includegraphics[width = \columnwidth]{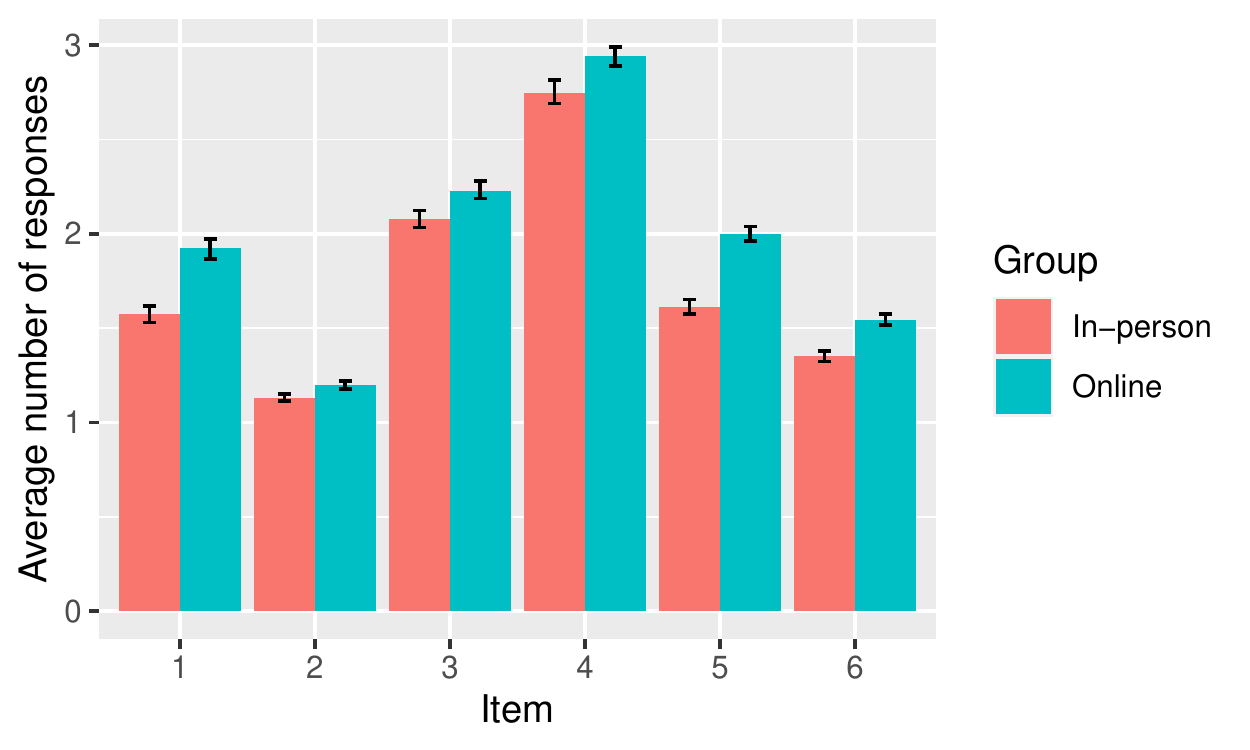}
    \caption{Average number of answer choices selected by Institution 1 students for the six MCMR PIQL items. Items 1, 2, 5, and 6 have one correct answer choice; item 3 has three correct answer choices, and 4 has two. Error bars indicate 95\% confidence intervals calculated using the bootstrap method \cite{tibshirani1993,davison2002,ggplot}.}
    \label{fig:avResp}
\end{figure}

\begin{figure}

    \includegraphics[width = \columnwidth]{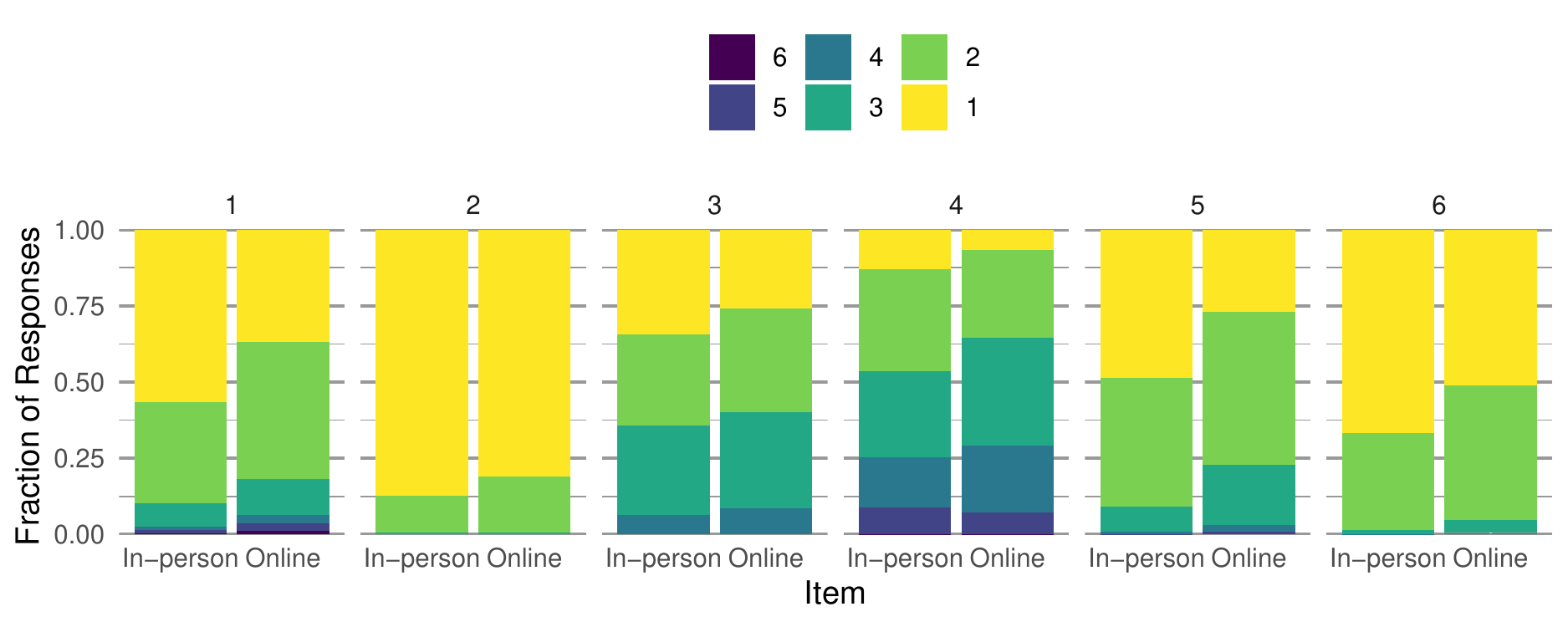}
    \caption{Fraction of student responses that include 1--7 answer choices for each item. Items 1 and 6 have six answer choices; items 2, 3, and 5 have five answer choices; and item 4 has seven answer choices. Items 1, 2, 5, and 6 have one correct answer choice; items 3 and 4 each have two. }
    \label{fig:numR}
\end{figure}

To explore whether administration method (i.e., electronic or on-paper) has an effect on number of responses chosen, we compare data collected with the paper version to those collected using the online version of the PIQL. We calculated the average number of responses for each of the six MCMR items for both online ($N=1404$) and in-person (paper) ($N=1697$) administrations at Institution 1. The results are shown in Fig. \ref{fig:avResp}. For each of the six items, the difference in the average number of answer choices is statistically significant (Mann-Whitney U test, $p < .001$ for all items), with effect size ranging from small to medium (rank-biserial correlation coefficients between $-0.0799$ and $-0.258$). Fig. \ref{fig:numR} shows the distributions of number of answer choices for each of the six items and suggests that the differences in averages seem to be due to students choosing one or two more responses online compared to on paper.

\begin{figure}

    \includegraphics[width = \columnwidth]{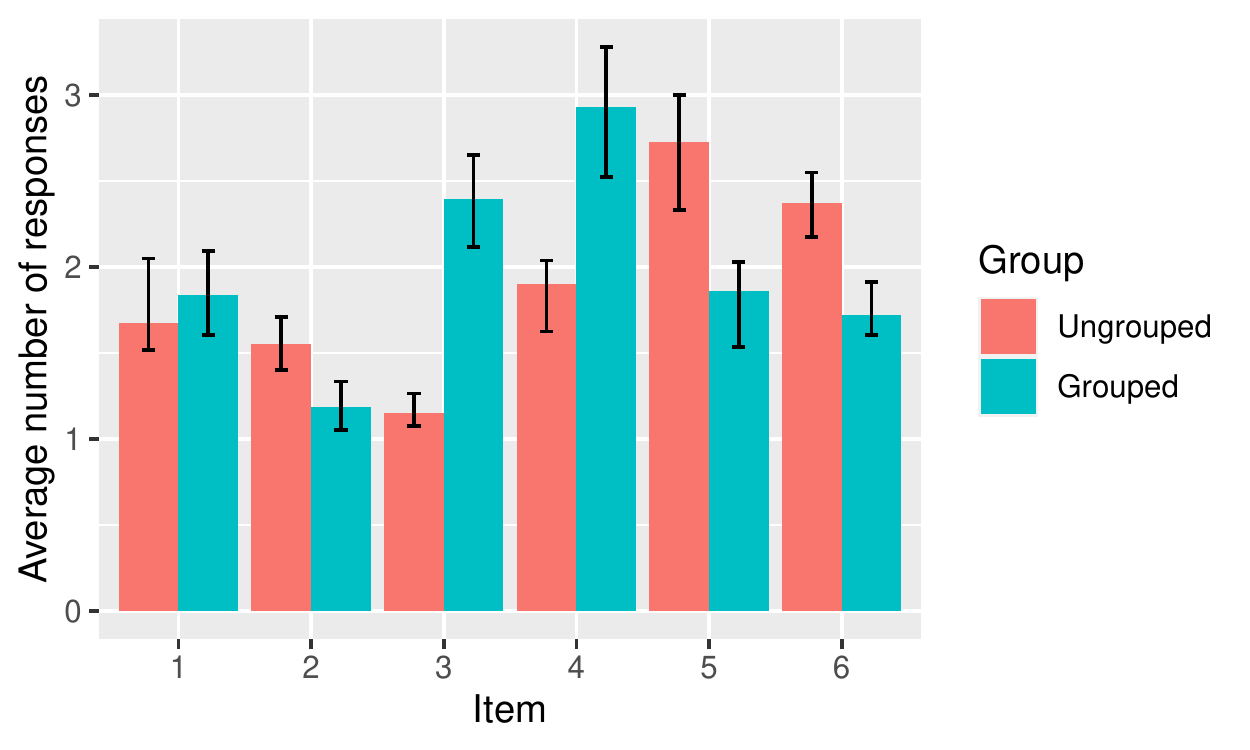}
    \caption{Average number of answer choices selected online by Institution 1 students for the six MCMR PIQL items with items grouped or ungrouped. Items 1, 2, 5, and 6 have one correct answer choice; item 3 has three correct answer choices, and 4 (the Work item shown in Fig. \ref{fig:workProb}) has two. Error bars indicate 95\% confidence intervals calculated using the bootstrap method \cite{tibshirani1993,davison2002,ggplot}.}
    \label{fig:avRespUG}
\end{figure}

To assess the effect of question order on the number of responses chosen, we compared the average number of answer choices for each of the MCMR items for students that saw the MCMR items grouped or ungrouped during online administration of the PIQL. Results are shown in Fig. \ref{fig:avRespUG}. While the number of answer choices seems to be statistically significantly different for all items except item 1, we note that there is no systematic effect.

\begin{figure}

    \includegraphics[width = \columnwidth]{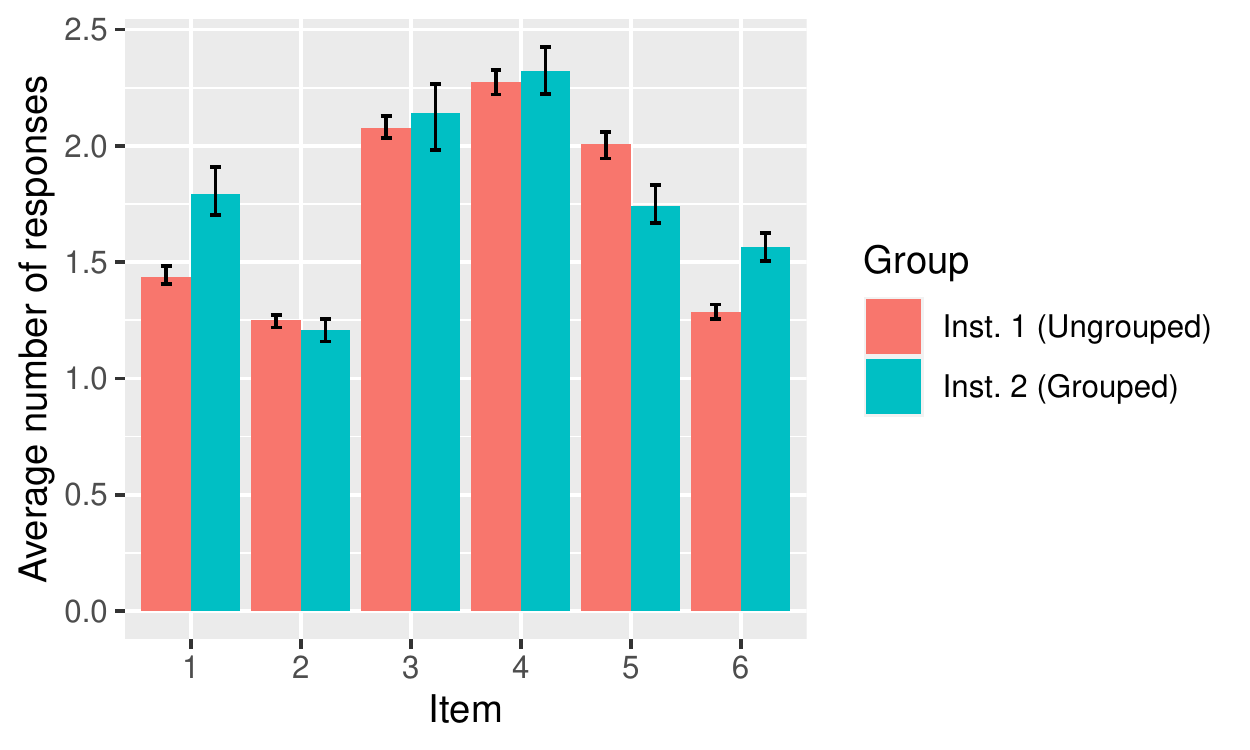}
    \caption{Average number of answer choices selected on paper by Institution 1 and Institution 2 students for the six MCMR PIQL items with items grouped or ungrouped. Items 1, 2, 5, and 6 have one correct answer choice; item 3 has three correct answer choices, and 4 has two. Error bars indicate 95\% confidence intervals calculated using the bootstrap method. \cite{tibshirani1993,davison2002,ggplot}.}
    \label{fig:avRespUGUWAF}
\end{figure}

Preliminary data collected at Institution 2 ($N=282$) and Institution 1 ($N=1404$) using paper versions of the PIQL in Fig. \ref{fig:avRespUGUWAF} also suggest that there is no systematic significant difference in the number of answer choices selected when items are grouped or ungrouped. We note, however, that comparing different populations of students can be difficult, as overall performance on the PIQL and MCMR response patterns between these populations differed. These results suggest that question order does not have a strong effect on number of answer choices selected, but also indicate more in-depth research would be appropriate.

\section{Discussion} 

For MCMR items, dichotomous scoring methods require a student to choose \emph{all} correct responses and \emph{only} correct responses to be considered correct. For example, MCMR item 4 on the PIQL (the Work item shown in Fig. \ref{fig:workProb}) has two correct answer choices: d and g. In a dichotomous scoring scheme a student who picks only answer d would be scored the same way as a student who chooses answers e and f (both incorrect). This ignores the nuance and complexity of students' response patterns within (and between) items. In an effort to move beyond the constraints of dichotomous scoring for MCMR items, we have developed a four-level scoring scale in which we categorize students' responses as Completely Correct, Some Correct (if at least one but not all correct response choices are chosen), Both Correct and Incorrect (if at least one correct and one incorrect response choices are chosen), and Completely Incorrect \cite{Smith2018, Smith2019rume}. 

\begin{figure}

    \includegraphics[width = \columnwidth]{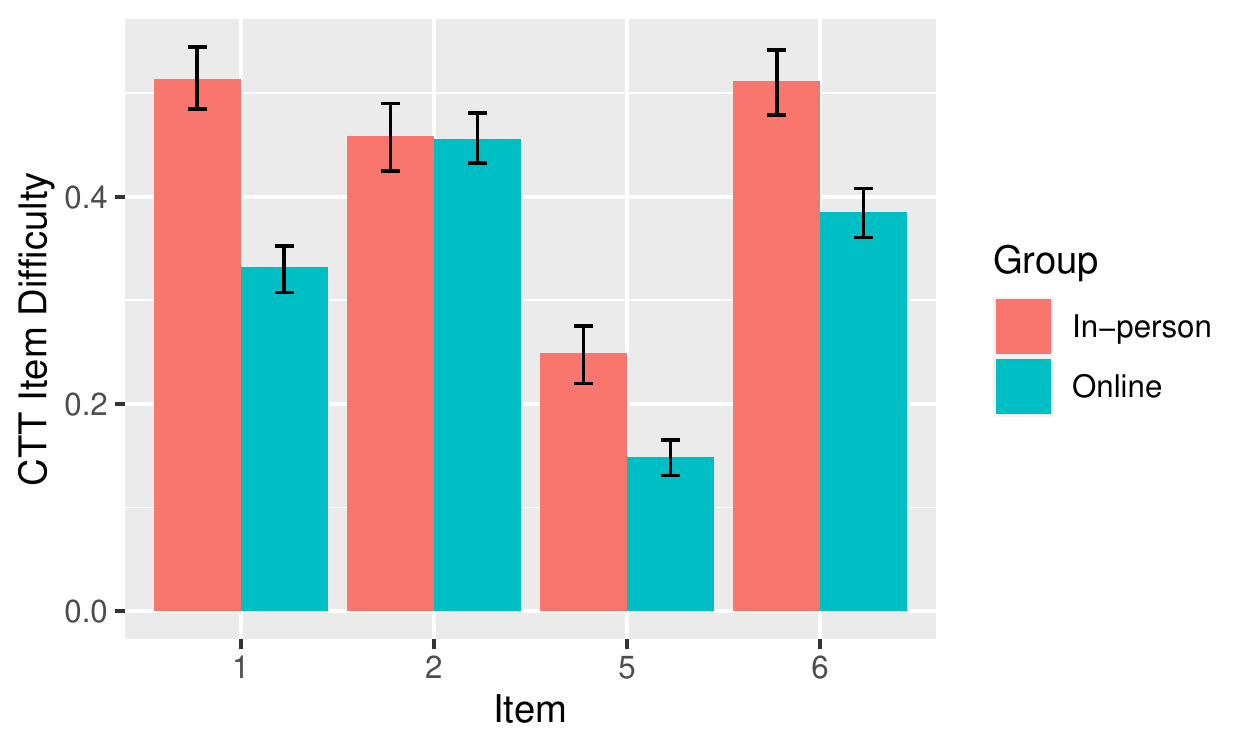}
    \caption{Item difficulty for the PIQL's four MCMR items with a single correct answer choice. Error bars indicate 95\% confidence intervals calculated using the bootstrap method \cite{tibshirani1993,davison2002,ggplot}.}
    \label{fig:difficulty}
\end{figure}

\begin{figure}

    \includegraphics[width = \columnwidth]{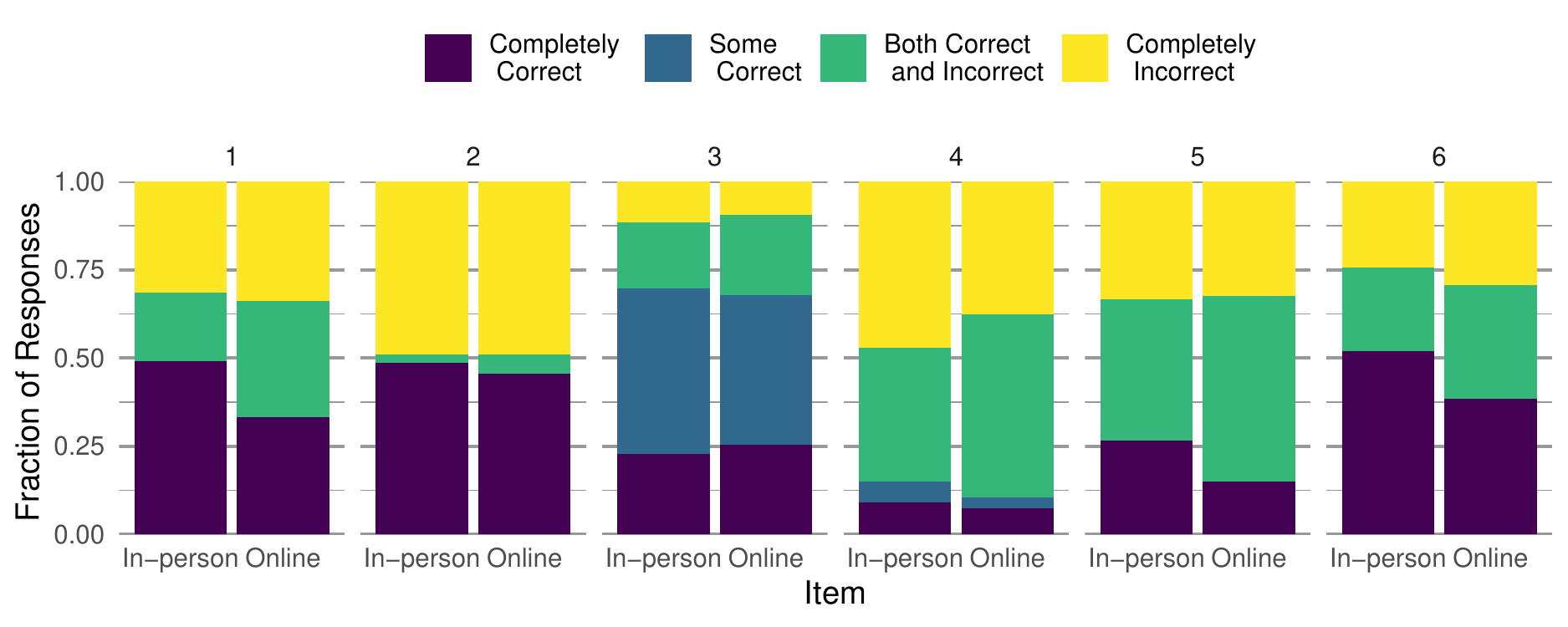}
    \caption{Fraction of student responses in each category of our four-level scoring scheme for MCMR items. We note that items 1, 2, 5, and 6 have one correct answer choice; items 3 and 4 each have two. }
    \label{fig:correctness}
\end{figure}

Only two of the MCMR items on the PIQL have more than one correct response. Therefore, an increase in the number of answers chosen is not necessarily associated with an improvement in performance. Indeed, for all four MCMR items with a single correct answer choice, student performance decreased substantially when the items were administered online and scored using dichotomous scoring methods. Fig. \ref{fig:difficulty} shows how the classical test theory (CTT) difficulty changes with administration method for the PIQL's four MCMR items with a single response. (CTT difficulty is the percentage of students answering completely correctly; therefore, a decrease in an item's difficulty is associated with a decrease in student performance on that item.) For three of the four items (1, 5, and 6), the difference in difficulty is statistically significant (binomial test $p < .001$), though the effect size is small (Cohen's $h$, $0.38<h<0.46$). To investigate more thoroughly how the increase in the number of answer choices associated with online administration was affecting student performance, we used the four-level scoring scale described above. The results are shown in Fig. \ref{fig:correctness}. We note that the dark purple ``completely correct'' bars in Fig. \ref{fig:correctness} represent the percentage of students that would be scored as answering a given item correctly if a dichotomous scoring method is used.

The results shown in Fig. \ref{fig:correctness} suggest that the percentage of students choosing no correct responses (the yellow bars) does not change substantially when the PIQL's MCMR items are administered online. Decreases in CTT item difficulty (i.e., the percentage of students answering completely correctly when a dichotomous scoring method is used) are instead associated with students choosing incorrect responses in addition to correct responses. This effect is particularly apparent for items 1, 4, 5, and 6. All of these items focus on reasoning about the meaning of sign associated with various quantities or quantitative relationships. Items 2 and 3, for which this effect is less pronounced, focus on proportions and scaling---while quantitative reasoning is required for these items, the answer choices themselves do not ask students to identify correct reasoning. PIQL items about sign and signed quantities generally have answer choices that include explicit reasoning or interpretation, whereas PIQL items to assess proportional and covariational reasoning typically do not include explanations in the answer choices. Thus it is difficult to tell whether the increase in answer choices following this pattern is associated with the topic (sign and signed quantities) or answer choice type (explicit reasoning); however, it seems unlikely that the effect is limited to items that probe student reasoning about sign.

\section{Conclusions}

The data presented in this paper suggest that students choose more responses to PIQL MCMR items when those items are administered online rather than on paper. This effect seems to be more pronounced for MCMR items that include explicit reasoning or interpretation in the answer choices. We conclude that online and on-paper administration methods of MCMR items provide different pictures of student reasoning, though we have not yet characterized this difference. It is possible that the ease of choosing multiple responses on the online version of the assessment is a ``nudge'' that inspires students to choose more responses that more completely represents their reasoning about a physics context, or perhaps encourages students to choose responses in which they have less confidence \cite{thalernudge}. The effect may be due to a combination of these reasons, or to other reasons we have not yet considered. 

Although additional analyses of student response patterns on the MCMR items from online and in-person administrations of the PIQL may provide some insight about the reasons for the observed differences, we expect that student interviews will be necessary to understand how students are interacting with MCMR items online. Data collected from such interviews could be compared to existing data from interviews in which MCMR items were presented on paper.  

Instructors should realize that responses collected via online administration of MCMR items may not be comparable to those collected in-person. Our data indicate student performance on MCMR items is likely to decrease with online administration when the items are scored using dichotomous scoring methods; this decrease may not necessarily be associated with a difference in understanding of content. Also, when developing scoring methods beyond dichotomous scoring, it may be necessary to take the increased number of responses into account. These results indicate that even MCMR items with evidence of validity when administered in-person should be subjected to validity checks for online use. 

Finally, we suggest that MCMR items provide a unique opportunity to explore how assessment differs online and in-person. We note there were no significant differences in student performance on the PIQL's SR items when we moved from on-paper to online administration \cite{white2021}; however, differences in performance on the PIQL's MCMR items suggest that administration method has an effect on how students interact with an assessment. Further study of MCMR items may help elucidate this effect.

\acknowledgements{This work is supported by the National Science Foundation under grants No. DUE-1832836, DUE- 1832880, DUE-1833050, DGE-1762114. Some of the work described in this paper was performed while the first author held an NRC Research Associateship award at Air Force Research Laboratory.}

\bibliography{MCMRbib.bib}

\end{document}